\documentclass[prd,reprint,onecolumn,notitlepage,preprintnumbers,superscriptaddress]{revtex4}
\usepackage{wrapfig}
\usepackage{amsfonts}
\usepackage{amsmath}
\usepackage{hyperref}
\usepackage{amssymb}
\usepackage[english]{babel}
\usepackage{graphicx}
\usepackage{epsfig}
\usepackage{bm}
\usepackage{longtable}
\usepackage{verbatim}
\usepackage{longtable}
\usepackage{color}
\usepackage{subfigure}
\usepackage[utf8]{inputenc}

\newcommand{\mathsym}[1]{{}}

\input{tcilatex}
\begin{document}

\title{A novel Randall-Sundrum model with $S_{3}$ flavor symmetry}
\author{A. E. C\'arcamo Hern\'andez}
\email{antonio.carcamo@usm.cl}
\affiliation{{\small Universidad T\'ecnica Federico Santa Mar\'{\i}a and Centro Cient%
\'{\i}fico-Tecnol\'ogico de Valpara\'{\i}so}\\
Casilla 110-V, Valpara\'{\i}so, Chile}
\author{I. de Medeiros Varzielas}
\email{ivo.de@soton.ac.uk}
\affiliation{{\small School of Physics and Astronomy, University of Southampton,}\\
Southampton, SO17 1BJ, U.K.}
\author{Nicol\'as A. Neill}
\email{naneill@ucdavis.edu}
\affiliation{{\small Physics Department, University of California, Davis}\\
Davis, California 95616, USA}

\begin{abstract}
We propose a simple and predictive model of fermion masses and mixing in a
warped extra dimension, with the smallest discrete non-Abelian group $S_{3}$
and the discrete symmetries $Z_{2}\otimes Z_{4}$. \ Standard Model fields
propagate in the bulk and the mass hierarchies and mixing angles are
accounted for the fermion zero modes localization profiles, similarly to the
the Randall-Sundrum (RS) model. To the best of our knowledge, this model is
the first implementation of an $S_{3}$ flavor symmetry in this type of
warped extra dimension framework. Our model successfully describes the
fermion masses and mixing pattern and is consistent with the current low
energy fermion flavor data. The discrete flavor symmetry in our model leads
to predictive mixing inspired textures, where the Cabbibo mixing arises from
the down type quark sector whereas up type quark sector contributes to the
remaining mixing angles.
\end{abstract}

\maketitle

\section{Introduction}

The recent LHC discovery of a $126$ GeV Higgs boson \cite%
{Aad:2012tfa,Chatrchyan:2012xdj} confirms the great success of the Standard
Model (SM) in describing electroweak phenomena, but the SM nevertheless
remains with many unanswered issues \cite{Agashe:2014kda}. One of them is
the hierarchy problem that arises from the quadratic divergence of the Higgs
mass, suggesting the presence of some underlying physics in the gauge
symmetry breaking mechanism that is so far unknown. Another issue is that
the SM does not specify the Yukawa structures, has no justification for the
number of generations, and lacks an explanation for the large hierarchy of
the fermion masses, spanning 5 orders of magnitude in the quark sector and a
much wider range when neutrinos are included. The origin of fermion mixing
and the size of CP violation in the quark and lepton sector is also a
related issue. While the mixing angles in the quark sector are very small,
in the lepton sector two of the mixing angles are large, and one mixing
angle is small. Neutrino experiments have brought clear evidence of neutrino
oscillations from the measured neutrino mass squared splittings. The three
neutrino flavors mix and at least two of the neutrinos have non vanishing
masses, which according to observations must be smaller than the SM charged
fermion masses by many orders of magnitude.

This flavor puzzle, not addressed in the framework of the SM, gives
motivation for extensions of the SM that explain the observed fermion mass
spectrum and flavor mixings. With neutrino experiments increasingly
constraining the mixing angles in the leptonic sector many models focus only
on this sector, aiming to explain the near tri-bi-maximal structure of the Pontecorvo-Maki-
Nakagawa-Sakata (PMNS) matrix through some non-Abelian symmetry. Furthermore, the fermion mass
hierarchy can be described by assuming textures for the Yukawa matrices, as
shown in Refs. \cite%
{Fritzsch:1977za,Fukuyama:1997ky,Du:1992iy,Barbieri:1994kw,Peccei:1995fg,Fritzsch:1999ee,Roberts:2001zy,Nishiura:2002ei,deMedeirosVarzielas:2005ax,Carcamo:2006dp,Kajiyama:2007gx,CarcamoHernandez:2010im,Branco:2010tx,Leser:2011fz,Gupta:2012dma,Hernandez:2013mcf,Pas:2014bra,Hernandez:2014hka,Hernandez:2014zsa,Nishiura:2014psa,Frank:2014aca,Sinha:2015ooa,Nishiura:2015qia,Gautam:2015kya,Pas:2015hca,Giraldo:2015cpp}%
. Discrete flavor groups implemented in several models provide a very
promising framework for describing the observed fermion mass and mixing
hierarchy (recent reviews on discrete flavor groups can be found in Refs. 
\cite{Ishimori:2010au,Altarelli:2010gt,King:2013eh, King:2014nza}). The
groups employed are quite diverse, mostly discrete subgroups of $SU(3)$ with
triplet representations. Of note are the groups $T_7$ \cite%
{Luhn:2007sy,Hagedorn:2008bc,Cao:2010mp,Luhn:2012bc,Kajiyama:2013lja,Bonilla:2014xla,Hernandez:2015cra,Arbelaez:2015toa,Hernandez:2015yxx}
and $\Delta(27)$ \cite%
{deMedeirosVarzielas:2006fc,Ma:2006ip,Varzielas:2012nn,Bhattacharyya:2012pi,Ma:2013xqa,Nishi:2013jqa,Varzielas:2013sla,Aranda:2013gga,Varzielas:2013eta,Abbas:2014ewa,Varzielas:2015aua,Abbas:2015zna}
as the smallest with distinct triplet and anti-triplet representations. $A_4$%
, the smallest group with a triplet representation \cite%
{Ma:2001dn,Babu:2002dz,Altarelli:2005yp,Altarelli:2005yx,deMedeirosVarzielas:2005qg,He:2006dk}
was one of the first groups explored and remains very popular after the
measurement of $\theta_{13}$ \cite%
{Varzielas:2012ai,Ishimori:2012fg,Ahn:2013mva,Memenga:2013vc,Bhattacharya:2013mpa,Ferreira:2013oga,Felipe:2013vwa,Hernandez:2013dta,King:2013hj,Morisi:2013qna,Morisi:2013eca,Felipe:2013ie,Campos:2014lla,Hernandez:2015tna,Karmakar:2015jza,Pramanick:2015qga}%
, as does the group $S_{4}$, which also has triplet representations \cite%
{Mohapatra:2012tb,Varzielas:2012pa,Ding:2013hpa,Ishimori:2010fs,Ding:2013eca,Hagedorn:2011un,Campos:2014zaa}%
.

Although it does not have a triplet irreducible representation, $S_3$ has
been considerably studied in the literature. It is appealing as it is the
smallest non-Abelian group, and can lead to interesting mass structures for
quarks, leptons or both. It was used as a flavor symmetry for the first time
by \cite{Pakvasa:1977in} and continues to be explored since $\theta_{13}$
was measured \cite%
{Cardenas:2012bg,Dias:2012bh,Dev:2012ns,Meloni:2012ci,Canales:2013cga,Ma:2013zca,Kajiyama:2013sza,Hernandez:2013hea,Hernandez:2014lpa,Hernandez:2014vta,Vien:2014vka,Ma:2014qra,Das:2015sca,Hernandez:2015dga}%
.

A scenario that remains relatively unexplored is of combining discrete
flavor symmetries with extra dimensions \cite%
{Altarelli:2005yp,Csaki:2008qq,Chen:2009gy,Kadosh:2010rm,Ishimori:2010fs,Kadosh:2011id,Kadosh:2013nra,Chen:2015jta}%
. In this work we consider using $S_{3}$ in a modified Randall-Sundrum 1
(RS1) \cite{Randall:1999ee} model, where the Higgs doublet is in the bulk.
In this implementation, the warped extra dimension controls the hierarchy
problem between the Electroweak (EW) scale and the Planck scale, and the
hierarchy in fermion masses, while the $S_{3}$ flavor symmetry is
responsible for making the flavor sector more predictive. This paper is
organized as follows. In section \ref{model} we outline the proposed model.
In section \ref{quarksector} we discuss the implications of our model in
quark masses and mixings. In section \ref{leptonsector} we present our
results on lepton masses and mixings. We conclude in section \ref%
{conclusions}. In Appendix \ref{S3} we present a brief description of the $%
S_3$ discrete group.

\section{The model}

\label{model} We consider the RS1 model based on a warped extra dimension
compactified on an ${S^{1}}/{Z_{2}}$ orbifold, which corresponds to the
interval $\left[ 0,\pi R\right] $, and with a metric of anti-de Sitter (AdS)
type given by \cite{Randall:1999ee}: 
\begin{equation}
ds^{2}=e^{-2ky}\eta _{\mu \nu }dx^{\mu }dx^{\nu }+dy^{2}.
\end{equation}

Here $\eta _{\mu \nu }=diag\left( -1,1,1,1\right) $ is the four-dimensional
Minkowski metric and $k$ the AdS curvature. Since the Planck mass $M_{P}\sim
10^{19}$ GeV is the fundamental scale, a natural theory should have $k\sim
M_{P}$. The TeV scale can be generated at the brane located at $y=\pi R$ if
the compactification radius $R$ is such that $ke^{-\pi kR}=1$ TeV, which in
turn means $kR\simeq 12$, also a rather natural number. The electroweak
gauge symmetry in the bulk has to be extended to $SU\left( 2\right)
_{L}\otimes SU\left( 2\right) _{R}\otimes U\left( 1\right) _{B-L}$ in order
to avoid a too large violation of the custodial symmetry caused by $U(1)_{Y}$
Kaluza-Klein (KK) modes \cite%
{Gherghetta:2000qt,Agashe:2003zs,Agashe:2006at,Csaki:2003zu,Cacciapaglia:2004rb,Csaki:2008qq,Gherghetta:2010cj,Alvarado:2012vn,Ponton:2012bi,CarcamoHernandez:2012xy,Frank:2015sua}. Since the custodial symmetry is preserved in our model, the $T$ parameter is consistent with the experimental data. Due to the preserved custodial symmetry in the bulk, the lower bound on the KK masses will mainly arise from the $S$ parameter constraint and is close to about $3$ TeV \cite{Agashe:2003zs,Csaki:2008qq}. An alternative approach to suppress contributions to the $T$ parameter is to consider a modified class of metrics that depart from AdS in the IR region \cite{Cabrer:2010si}. 

Our model is an extension of the SM with seven EW scalar singlets and two
heavy Majorana neutrinos, embedded in a warped extra dimension, with the
inclusion of the $S_{3}$ discrete flavor symmetry. The inclusion of the
discrete $S_{3}$ symmetry serves to reduce the number of parameters in the
Yukawa sector of our model making it more predictive. We assume that 3 EW
scalar singlets and two heavy Majorana neutrinos are located at the Planck
brane, whereas the remaining EW scalar singlets are set at the TeV brane. The Higgs doublet as well as the three generations
of fermions propagate in the 5-dimensional bulk.


Boundary conditions at the $y=0$ and $y=\pi R$ branes determine a tower of
KK modes, of which the zero modes are assumed to be the SM fields in 4-D.
Since there is no chirality in 5 dimensions, the left- and right-handed
fermions in 4-D correspond to zero modes of different fields in the bulk,
conventionally called $\Psi _{R}$ and $\Psi _{L}$, which obey the conditions 
$\Psi _{R,L}=\pm \gamma _{5}\Psi _{R,L}$. The masses of the zero modes in
4-D appear when the electroweak symmetry is spontaneously broken, as in the
SM. Fermion fields in the bulk of 5D space, generically denoted as $\Psi
(x,y)$, obey an action of the form: 

\begin{equation}
S_{5}^{(\Psi )}=\int d^{4}x\int_{0}^{\pi R}dy\sqrt{-g}\left( i\overline{\Psi 
}\Gamma ^{M}\nabla _{M}\Psi +im_{\Psi }\overline{\Psi }\Psi +\frac{\lambda
_{ij}}{M_{5}^{\frac{1}{2}}}\overline{\Psi }_{L}H\Psi _{R}\right) ,
\label{SF}
\end{equation}

where capital indices run over the five coordinates, $\Gamma
^{M}=(e^{ky}\gamma ^{\mu },\gamma ^{5})$ are the Gamma matrices in the 5D
curved spacetime and $\nabla _{M}$ is the covariant derivative that includes
the interaction with the gauge fields. The fermion masses in 5-D have
natural values of the order of the Planck mass: 
\begin{equation}
m_{\Psi }=k\ d_{\Psi },  \label{localizations}
\end{equation}%
where $k$ is the AdS curvature and $d_{\Psi }$ are parameters of order
unity. These parameters will determine the localization of the fermion
profiles along the 5th dimension. The masses of fermions in the TeV brane
(the observable 4D space) will be a consequence of this localization after
electroweak symmetry breaking. $M_{5}$ is the 5-D Planck mass, which is
related to $M_{P}$, the Planck mass in 4-D, through the following relation: 
\begin{equation}
M_{P}^{2}=\frac{M_{5}^{3}}{k}\left( 1-e^{-2\pi kR}\right) .
\end{equation}%
Since the exponential term is highly suppressed, $M_{5}$, $M_{P}$ and $k$
are all of the same order of magnitude. %
%
%
%
%
%
%

Fermion, gauge boson fields and scalars can be expanded in their respective
KK modes as follows: 
\begin{eqnarray}
\Psi _{L,R}\left( x,y\right) &=&\sum_{n=0}^{\infty }\Psi _{L,R}^{\left(
n\right) }\left( x\right) f_{L,R}^{\left( n\right) }\left( y\right) , \\
A_{\mu }^{a}\left( x,y\right) &=&\sum_{n=0}^{\infty }A_{\mu }^{a\left(
n\right) }\left( x\right) \chi ^{\left( n\right) }\left( y\right) , \\
H\left( x,y\right) &=&\sum_{n=0}^{\infty }H_{\mu }^{\left( n\right) }\left(
x\right) h^{\left( n\right) }\left( y\right) ,
\end{eqnarray}%
where their profiles in the 5th dimension are respectively given by \cite%
{Gherghetta:2010cj}: 
\begin{equation}
f_{L,R}^{\left( n\right) }\left( y\right) =\left\{ 
\begin{array}{ll}
\sqrt{\frac{k\left( 1-2d_{L,R}\right) }{e^{\left( 1-2d_{L,R}\right) k\pi R}-1%
}}e^{\left( 2-d_{L,R}\right) ky} & ,\ n=0 \\ 
N_{\Psi _{L,R}}^{\left( n\right) }e^{\frac{5}{2}ky}\left[ J_{d_{L,R}\pm 
\frac{1}{2}}\left( \frac{m_{\Psi }^{\left( n\right) }}{k}e^{ky}\right)
+\alpha _{\Psi _{L,R}}^{\left( n\right) }Y_{d_{L,R}\pm \frac{1}{2}}\left( 
\frac{m_{\Psi }^{\left( n\right) }}{k}e^{ky}\right) \right] & ,\ n>0,%
\end{array}%
\right.
\end{equation}%
\begin{equation}
\chi ^{\left( n\right) }\left( y\right) =\left\{ 
\begin{array}{ll}
\frac{1}{\sqrt{\pi R}} & ,\ n=0 \\ 
N_{A}^{\left( n\right) }e^{ky}\left[ J_{1}\left( \frac{m_{A}^{\left(
n\right) }}{k}e^{ky}\right) +\alpha _{A}^{\left( n\right) }Y_{1}\left( \frac{%
m_{A}^{\left( n\right) }}{k}e^{ky}\right) \right] & ,\ n>0.%
\end{array}%
\right.
\end{equation}

\begin{equation}
h^{\left( n\right) }\left( y\right) =\left\{ 
\begin{array}{ll}
\sqrt{\frac{2k\left( 1\pm \sqrt{4+a}\right) }{\left[ e^{2\left( 1\pm \sqrt{%
4+a}\right) k\pi R}-1\right] }}e^{\left( 2\pm \sqrt{4+a}\right) ky} & ,\ n=0
\\ 
N_{H}^{\left( n\right) }e^{2ky}\left[ J_{\pm \sqrt{4+a}}\left( \frac{%
m_{H}^{\left( n\right) }}{k}e^{ky}\right) +\alpha _{h}^{\left( n\right)
}Y_{\pm \sqrt{4+a}}\left( \frac{m_{H}^{\left( n\right) }}{k}e^{ky}\right) %
\right] & ,\ n>0.%
\end{array}%
\right.
\end{equation}%
Here $\chi ^{\left( n\right) }\left( y\right) $ are given for the gauge
where $A_{5}=0$. 
The functions $J_{\rho }$, $Y_{\rho }$ are first and second kind Bessel
functions, respectively and $N_{\Psi _{L,R}}^{\left( n\right) }$, $%
N_{A}^{\left( n\right) }$ are normalization constants computed from the
following orthonormality relations: 
\begin{equation}
\int_{0}^{\pi R}dye^{-3ky}f_{L,R}^{\left( n\right) }\left( y\right)
f_{L,R}^{\left( m\right) }\left( y\right) =\delta _{nm},\hspace{0.7cm}%
\int_{0}^{\pi R}dy\chi ^{\left( n\right) }\left( y\right) \chi ^{\left(
m\right) }\left( y\right) =\delta _{nm},\hspace{0.7cm}\int_{0}^{\pi
R}dye^{-2ky}h^{\left( n\right) }\left( y\right) h^{\left( m\right) }\left(
y\right) =\delta _{nm}.
\end{equation}%
The coefficients $\alpha _{\Psi _{L,R}}^{\left( n\right) }$, $\alpha
_{A}^{\left( n\right) }$ and $\alpha _{H}^{\left( n\right) }$ are determined
by the boundary conditions on the branes, resulting in the following
relations: 
\begin{eqnarray}
\alpha _{\Psi _{L,R}}^{\left( n\right) } &=&-\frac{J_{\pm d_{L,R}\pm \frac{1%
}{2}}\left( \frac{m_{\Psi }^{\left( n\right) }}{k}\right) }{Y_{\pm
d_{L,R}\pm \frac{1}{2}}\left( \frac{m_{\Psi }^{\left( n\right) }}{k}\right) }%
=-\frac{J_{\pm d_{L,R}\pm \frac{1}{2}}\left( \frac{m_{\Psi }^{\left(
n\right) }}{k}e^{\pi kR}\right) }{Y_{\pm d_{L,R}\pm \frac{1}{2}}\left( \frac{%
m_{\Psi }^{\left( n\right) }}{k}e^{\pi kR}\right) }, \\
\alpha _{A}^{\left( n\right) } &=&-\frac{J_{0}\left( \frac{m_{A}^{\left(
n\right) }}{k}\right) }{Y_{0}\left( \frac{m_{A}^{\left( n\right) }}{k}%
\right) }\quad =-\frac{J_{0}\left( \frac{m_{A}^{\left( n\right) }}{k}e^{\pi
kR}\right) }{Y_{0}\left( \frac{m_{A}^{\left( n\right) }}{k}e^{\pi kR}\right) 
},  \label{alphaA} \\
\alpha _{H}^{\left( n\right) } &=&-\frac{J_{0}\left( \frac{m_{H}^{\left(
n\right) }}{k}\right) }{Y_{0}\left( \frac{m_{H}^{\left( n\right) }}{k}%
\right) }\quad =-\frac{J_{0}\left( \frac{m_{H}^{\left( n\right) }}{k}e^{\pi
kR}\right) }{Y_{0}\left( \frac{m_{H}^{\left( n\right) }}{k}e^{\pi kR}\right) 
}.
\end{eqnarray}%
In turn, these relations determine the masses $m_{\Psi }^{\left( n\right) }$%
, $m_{A}^{\left( n\right) }$ and $m_{H}^{\left( n\right) }$ of the non-zero
modes. These modes correspond to heavy particles beyond the spectrum of the
SM. On the other hand, the zero modes, which are identified with the SM
fermions remain massless at this level and become massive only after
electroweak symmetry breaking.

Quark and lepton fields are assigned into two $S_{3}$ doublets and several $S_{3}$ singlets, as follows:

{%
\begin{eqnarray}
Q_{L} &=&\left( q_{1L},q_{2L}\right) \sim \mathbf{2},\hspace{1cm}q_{3L}\sim 
\mathbf{1}^{\prime },\hspace{1cm}u_{1R}\sim \mathbf{1},\hspace{1cm}%
u_{2R}\sim \mathbf{1}^{\prime },\hspace{1cm}u_{3R}\sim \mathbf{1}^{\prime },
\notag \\
d_{1R} &\sim &\mathbf{1},\hspace{1cm}d_{2R}\sim \mathbf{1,\hspace{1cm}}%
d_{3R}\sim \mathbf{1}^{\prime }\mathbf{,}  \label{Fermionassignments} \\
l_{R} &=&\left( l_{1R},l_{2R}\right) \sim \mathbf{2},\hspace{1cm}l_{3R}\sim 
\mathbf{1}^{\prime },\hspace{1cm}l_{1L}\sim \mathbf{1},\hspace{1cm}%
l_{2L}\sim \mathbf{1}^{\prime },\hspace{1cm}l_{3L}\sim \mathbf{1}^{\prime }%
\mathbf{,}  \notag \\
\nu _{1R} &\sim &\mathbf{1}^{\prime },\hspace{1cm}\nu _{2R}\sim \mathbf{1}%
^{\prime }.
\end{eqnarray}%
}

The $SU(2)_{L}$ scalar doublet $H$ is assigned as a trivial $S_{3}$ singlet
whereas the different EW scalar singlets are grouped into two $S_{3}$
doublets, one $S_{3}$ trivial singlet and two $S_{3}$ non-trivial singlets.
The scalar assignments under the $S_{3}$ flavor symmetry are:

\begin{eqnarray}
H &\sim &\mathbf{1},\hspace{1cm}\zeta \sim \mathbf{1}^{\prime },\hspace{1cm}%
\eta \sim \mathbf{1},\hspace{1cm}\rho \sim \mathbf{1}^{\prime }, \\
\chi &=&\left( \chi _{1},\chi _{2}\right) \sim \mathbf{2},\hspace{1cm}\xi
=\left( \xi _{1},\xi _{2}\right) \sim \mathbf{2.}
\end{eqnarray}

We assume that $\chi $ and $\zeta $ are located in the TeV brane whereas the
remaining scalar singlets, i.e., $\eta $ , $\rho $ and $\xi $\ are set at
the Planck brane.

The minimization equations for the scalar potentials for a single $S_{3}$
scalar doublet can be seen in Ref. \cite{Hernandez:2015dga}. With $\xi $ and 
$\chi $ being in different branes, their minimization can proceed
independently (similarly to the strategy of \cite{Altarelli:2005yp}). They
acquire the following vacuum expectation value (VEV) pattern: 
\begin{equation}
\left\langle \xi \right\rangle =v_{\xi }\left( 1,0\right) ,\hspace{1cm}%
\left\langle \chi \right\rangle =v_{\chi }\left( 1,0\right) ,
\end{equation}%
so that the VEVs are both pointing in the $\left( 1,0\right) $ $S_{3}$
direction.

Furthermore, we include a discrete $Z_{2}\otimes Z_{4}$ symmetry. The $Z_{2}$
symmetry decouples the bottom quark from the down and strange quarks. The
fields charged under $Z_{2}$ transform as follows:

\begin{eqnarray}
Q_{L} &\rightarrow &-Q_{L},\hspace{1cm}u_{1R}\rightarrow -u_{1R},\hspace{1cm}%
u_{2R}\rightarrow -u_{2R},\hspace{1cm}  \notag \\
d_{1R} &\rightarrow &-d_{1R},\hspace{1cm}d_{2R}\rightarrow -d_{2R},\hspace{%
1cm}\chi \rightarrow -\chi ,
\end{eqnarray}%
while the fields charged under $Z_{4}$ have the following non-trivial
transformation properties: 
\begin{eqnarray}
d_{1R} &\rightarrow &id_{1R},\hspace{1cm}d_{2R}\rightarrow -d_{2R},\hspace{%
1cm}d_{3R}\rightarrow id_{3R}, \\
\rho &\rightarrow &-\rho ,\hspace{1cm}\eta \rightarrow -i\eta ,\hspace{1cm}%
l_{1L}\rightarrow il_{1L}.
\end{eqnarray}

The $SU(2)_{L}\otimes SU(2)_{R}\otimes U(1)_{B-L}\otimes S_{3}\otimes
Z_{2}\otimes Z_{4}$ invariant quark and lepton Yukawa interactions are: 
\begin{eqnarray}
S_{5}^{(q)} &=&\int d^{4}x\int_{0}^{\pi R}dy\sqrt{-g}\left\{ \left[ \frac{%
\lambda _{13}^{\left( u\right) }}{M_{5}^{\frac{7}{2}}}\overline{Q}_{L}%
\widetilde{H}u_{3R}\chi \zeta +\frac{\lambda _{23}^{\left( u\right) }}{%
M_{5}^{\frac{5}{2}}}\overline{Q}_{L}\widetilde{H}u_{3R}\chi \right] \delta
\left( y-\pi R\right) +\frac{\lambda _{33}^{\left( u\right) }}{M_{5}^{\frac{1%
}{2}}}\overline{q}_{3L}\widetilde{H}u_{3R}\right\}  \notag \\
&&+\int d^{4}x\int_{0}^{\pi R}dy\sqrt{-g}\left[ \frac{\lambda _{11}^{\left(
u\right) }}{M_{5}^{\frac{5}{2}}}\overline{Q}_{L}\widetilde{H}u_{1R}\xi +%
\frac{\lambda _{22}^{\left( u\right) }}{M_{5}^{\frac{5}{2}}}\overline{Q}_{L}%
\widetilde{H}u_{2R}\xi +\frac{\lambda _{11}^{\left( d\right) }}{M_{5}^{\frac{%
7}{2}}}\overline{Q}_{L}Hd_{1R}\xi \eta \right.  \notag \\
&&+\left. \frac{\lambda _{22}^{\left( d\right) }}{M_{5}^{\frac{7}{2}}}%
\overline{Q}_{L}Hd_{2R}\xi \rho +\frac{\lambda _{12}^{\left( d\right) }}{%
M_{5}^{\frac{9}{2}}}\overline{Q}_{L}Hd_{2R}\xi \eta ^{2}+\frac{\lambda
_{33}^{\left( d\right) }}{M_{5}^{\frac{5}{2}}}\overline{q}_{3L}Hd_{3R}\eta %
\right] \delta \left( y\right),  \label{LYq}
\end{eqnarray}

\begin{eqnarray}
S_{5}^{(l)} &=&\int d^{4}x\int_{0}^{\pi R}dy\sqrt{-g}\left[ \frac{\lambda
_{11}^{\left( l\right) }}{M_{5}^{\frac{7}{2}}}\overline{l}_{1L}Hl_{R}\xi
\eta ^{\ast }+\frac{\lambda _{22}^{\left( l\right) }}{M_{5}^{\frac{5}{2}}}%
\overline{l}_{2L}Hl_{R}\xi +\frac{\lambda _{32}^{\left( l\right) }}{M_{5}^{%
\frac{5}{2}}}\overline{l}_{3L}Hl_{R}\xi \right] \delta \left( y\right) 
\notag \\
&&+\int d^{4}x\int_{0}^{\pi R}dy\sqrt{-g}\left[ \frac{\lambda _{23}^{\left(
l\right) }}{M_{5}^{\frac{1}{2}}}\overline{l}_{2L}Hl_{3R}+\frac{\lambda
_{33}^{\left( l\right) }}{M_{5}^{\frac{1}{2}}}\overline{l}_{3L}Hl_{3R}\right]
\notag \\
&&+\int d^{4}x\int_{0}^{\pi R}dy\sqrt{-g}\left[ \frac{\lambda _{32}^{\left(
\nu \right) }}{M_{5}}\overline{l}_{2L}\widetilde{H}\nu _{1R}+\frac{\lambda
_{32}^{\left( \nu \right) }}{M_{5}}\overline{l}_{3L}\widetilde{H}\nu _{1R}%
\right] \delta \left( y\right)  \notag \\
&&+\int d^{4}x\int_{0}^{\pi R}dy\sqrt{-g}\left[ \frac{\lambda _{11}^{\left(
\nu \right) }}{M_{5}^{2}}\overline{l}_{1L}\widetilde{H}\nu _{2R}\eta ^{\ast
}+\frac{\lambda _{21}^{\left( \nu \right) }}{M_{5}}\overline{l}_{2L}%
\widetilde{H}\nu _{2R}+\frac{\lambda _{22}^{\left( \nu \right) }}{M_{5}}%
\overline{l}_{3L}\widetilde{H}\nu _{2R}\right] \delta \left( y\right)  \notag
\\
&&+\int d^{4}x\int_{0}^{\pi R}dy\sqrt{-g}\left[ M_{1\nu }\overline{\nu }%
_{1R}\nu _{1R}^{c}+M_{2\nu }\overline{\nu }_{2R}\nu _{2R}^{c}+M_{3\nu
}\left( \overline{\nu }_{1R}\nu _{2R}^{c}+\overline{\nu }_{2R}\nu
_{1R}^{c}\right) \right] \delta \left( y\right) ,  \label{Lyl}
\end{eqnarray}

where the dimensionless couplings in Eqs. (\ref{LYq}) and (\ref{Lyl}) are $%
\mathcal{O}(1)$ parameters. Note that with $\chi $ in the TeV brane and $%
\eta $ in the Planck brane, terms like $\overline{Q}_{L}Hd_{3R}\chi \eta $
do not contribute and were therefore not included above, even though they
are invariant under all the symmetries.

Assuming that the quark mass and mixing pattern arises from the fermion
profiles along the extra dimension, we set the VEVs of the EW scalar
singlets with respect to the Wolfenstein parameter $\lambda =0.225$ and the
new physics scale $M_{5}$: 
\begin{equation}
v_{\zeta }\sim v_{\chi }\sim v_{\rho }\sim v_{\eta }\sim \lambda M_{5};\quad
v_\xi \sim \lambda^3 M_{5}.  \label{VEVsize}
\end{equation}

The presence of additional scalars (in the TeV brane) leads to the
interesting possibility of extra scalars at colliders. For recent studies on
Higgs production and decay in multi-Higgs models (some of which are embedded
into 5D warped models), see for instance Ref. \cite{Bhattacharyya:2010hp,Bhattacharyya:2012ze,
Malm:2013jia,Malm:2014gha,Archer:2014jca,Dillon:2014zea,Frank:2015zwd,Hernandez:2015dga}%
.

\section{Quark masses and mixings}

\label{quarksector} From the quark Yukawa interactions given by Eq. (\ref%
{LYq}) we get the following mass matrix textures for quarks

\begin{eqnarray}
M_{U} &=&\left( 
\begin{array}{ccc}
\varepsilon _{11}^{\left( u\right) }\frac{v_{\xi }}{M_{5}} & 0 & \varepsilon
_{13}^{\left( u\right) }\frac{v_{\chi }v_{\zeta }}{M_{5}^{2}} \\ 
0 & \varepsilon _{22}^{\left( u\right) }\frac{v_{\xi }}{M_{5}} & \varepsilon
_{23}^{\left( u\right) }\frac{v_{\chi }}{M_{5}} \\ 
0 & 0 & \varepsilon _{33}^{\left( u\right) }%
\end{array}%
\right) \frac{v}{\sqrt{2}},  \notag \\
M_{D} &=&\left( 
\begin{array}{ccc}
\varepsilon _{11}^{\left( d\right) }\frac{v_{\xi }v_{\eta }}{M_{5}^{2}} & 
\varepsilon _{12}^{\left( d\right) }\frac{v_{\xi }v_{\eta }^{2}}{M_{5}^{3}}
& 0 \\ 
0 & \varepsilon _{22}^{\left( d\right) }\frac{v_{\xi }v_{\rho }}{M_{5}^{2}}
& 0 \\ 
0 & 0 & \varepsilon _{33}^{\left( d\right) }\frac{v_{\eta }}{M_{5}}%
\end{array}%
\right) \frac{v}{\sqrt{2}},  \label{Mq}
\end{eqnarray}

where: 
\begin{eqnarray}
\varepsilon _{n3}^{\left( u\right) } &=&\lambda _{n3}^{\left( u\right) }%
\frac{k}{M_{5}^{\frac{3}{2}}}\sqrt{\frac{2k\left( 1\pm \sqrt{4+a}\right)
\left( 1-2y_{nL}^{\left( q\right) }\right) \left( 1-2y_{3R}^{\left( u\right)
}\right) }{\left[ e^{2\left( 1\pm \sqrt{4+a}\right) k\pi R}-1\right] \left[
e^{\left( 1-2y_{nL}^{\left( q\right) }\right) k\pi R}-1\right] \left[
e^{\left( 1-2y_{3R}^{\left( u\right) }\right) k\pi R}-1\right] }}e^{\left(
2-y_{nL}^{\left( q\right) }-y_{3R}^{\left( u\right) }\pm \sqrt{4+a}\right)
k\pi R},\hspace{1cm}n=1,2.  \notag \\
\varepsilon _{33}^{\left( u\right) } &=&\lambda _{33}^{\left( u\right) }%
\frac{1}{M_{5}^{\frac{1}{2}}}\sqrt{\frac{2k\left( 1\pm \sqrt{4+a}\right)
\left( 1-2y_{3L}^{\left( q\right) }\right) \left( 1-2y_{3R}^{\left( u\right)
}\right) }{\left[ e^{2\left( 1\pm \sqrt{4+a}\right) k\pi R}-1\right] \left[
e^{\left( 1-2y_{3L}^{\left( q\right) }\right) k\pi R}-1\right] \left[
e^{\left( 1-2y_{3R}^{\left( u\right) }\right) k\pi R}-1\right] }}\frac{%
e^{\left( 2-y_{3L}^{\left( q\right) }-y_{3R}^{\left( u\right) }\pm \sqrt{4+a}%
\right) k\pi R}-1}{\left[ 2-y_{3L}^{\left( q\right) }-y_{3R}^{\left(
u\right) }\pm \sqrt{4+a}\right] },  \notag \\
\varepsilon _{nn}^{\left( u\right) } &=&\lambda _{nn}^{\left( u\right) }%
\frac{k}{M_{5}^{\frac{3}{2}}}\sqrt{\frac{2k\left( 1\pm \sqrt{4+a}\right)
\left( 1-2y_{nL}^{\left( q\right) }\right) \left( 1-2y_{nR}^{\left( u\right)
}\right) }{\left[ e^{2\left( 1\pm \sqrt{4+a}\right) k\pi R}-1\right] \left[
e^{\left( 1-2y_{nL}^{\left( q\right) }\right) k\pi R}-1\right] \left[
e^{\left( 1-2y_{nR}^{\left( u\right) }\right) k\pi R}-1\right] }},\hspace{1cm%
}n=1,2.  \notag \\
\varepsilon _{ij}^{\left( d\right) } &=&\lambda _{ij}^{\left( d\right) }%
\frac{k}{M_{5}^{\frac{3}{2}}}\sqrt{\frac{2k\left( 1\pm \sqrt{4+a}\right)
\left( 1-2y_{iL}^{\left( q\right) }\right) \left( 1-2y_{jR}^{\left( u\right)
}\right) }{\left[ e^{2\left( 1\pm \sqrt{4+a}\right) k\pi R}-1\right] \left[
e^{\left( 1-2y_{iL}^{\left( q\right) }\right) k\pi R}-1\right] \left[
e^{\left( 1-2y_{jR}^{\left( d\right) }\right) k\pi R}-1\right] }},\hspace{1cm%
}i,j=1,2,3.  \label{lambda}
\end{eqnarray}

where $y_{1L}=y_{2L}=y_{L}$, as follows from the fact that $q_{1L}$ and $%
q_{2L} $ are unified into a $S_{3}$ doublet as seen from Eq. (\ref%
{Fermionassignments}).

We assume that the theory conserves CP and that CP is violated spontaneously
only through the complex VEV $v_{\zeta }${\ of the }$\zeta ${\ scalar. This
makes the model more predictive.} We fit the parameters in Eq. (\ref{Mq}) to
reproduce the quark masses and quark mixing parameters.

\bigskip

\begin{table}[tbh]
\begin{center}
\begin{tabular}{c|l|l}
\hline\hline
Observable & Model value & Experimental Value \\ \hline
$m_{u}(MeV)$ & \quad $1.48$ & \quad $1.45_{-0.45}^{+0.56}$ \\ \hline
$m_{c}(MeV)$ & \quad $634$ & \quad $635\pm 86$ \\ \hline
$m_{t}(GeV)$ & \quad $172.1$ & \quad $172.1\pm 0.6\pm 0.9$ \\ \hline
$m_{d}(MeV)$ & \quad $3.1$ & \quad $2.9_{-0.4}^{+0.5}$ \\ \hline
$m_{s}(MeV)$ & \quad $56.3$ & \quad $57.7_{-15.7}^{+16.8}$ \\ \hline
$m_{b}(GeV)$ & \quad $2.82$ & \quad $2.82_{-0.04}^{+0.09}$ \\ \hline
\end{tabular}%
\end{center}
\caption{Model and experimental values of quark masses.}
\label{Observables0}
\end{table}

In order to reproduce the experimental values of the 10 physical observables
of the quark sector, i.e., the 6 quark masses and 4 quark mixing parameters,
we fix the right-handed top quark localization profile and the Higgs
profile, as follows, $y_{1R}^{(u)}=y_{3R}^{(u)}$, $a=-3$, $\lambda_{ij}^{\left( u\right) }=\lambda_{ij}^{\left( d\right) }=10$,
whereas we fit the
effective 9 free parameters, i.e., the 2 left-handed quark profiles, the
remaining 5 right-handed quark profiles, and the magnitude and phase of $%
v_{\zeta }$. The values of the quark profiles, the magnitude and phase of $%
v_{\zeta }$, corresponding to the results reported in Tables \ref%
{Observables0} and \ref{obs} are
$y_{L}^{\left( q\right) }=0.502177$,
$y_{3L}^{(u)}=0.522259$,
$y_{1R}^{(u)}=0.41452$,
$y_{2R}^{(u)}=28.5089$,
$y_{1R}^{(d)}=0.500942$,
$y_{2R}^{(d)}=4.63314$,
$y_{3R}^{(d)}=0.532187$,
$|v_{\zeta }| = 0.4 \lambda M_5$, and $\alpha =67.6164^{\circ }$ (being $%
\alpha $ the complex phase of $v_{\zeta }$). Note that, as all
the left handed profiles profiles are $>1/2$, the flavor changing neutral current contributions due to KK modes are
suppressed \cite{Gherghetta:2000qt,Gherghetta:2010cj}. This suppression is due to the fact that the gauge coupling of zero mode fermions to Kaluza-Klein gauge bosons is much smaller than the electroweak coupling constant when the left handed fermion profiles are equal or larger than $1/2$.

The experimental values of the CKM matrix are taken from Ref.~\cite{Agashe:2014kda}. 
\begin{table}[tbh]
\begin{center}
\begin{tabular}{c|l|l}
\hline\hline
Observable & Model value & Experimental value \\ \hline
$\sin \theta _{12}$ & \quad $0.22$ & \quad $0.22536\pm 0.00061$ \\ \hline
$\sin \theta _{23}$ & \quad $0.0413$ & \quad $0.0414\pm 0.0012$ \\ \hline
$\sin \theta _{13}$ & \quad $0.0037$ & \quad $0.00355\pm 0.00015$ \\ \hline
$\delta $ & \quad $68^{\circ }$ & \quad $68^{\circ }$ \\ \hline\hline
\end{tabular}%
\end{center}
\caption{Model and experimental values of CKM parameters.}
\label{obs}
\end{table}
As can be seen, the quark masses and the Cabibbo-Kobayashi-Maskawa (CKM) matrix obtained from these
textures are in good agreement with the experimental data. The agreement of
our model with the experimental data is as good as in the models of Refs. 
\cite{Branco:2010tx,King:2013hj,Hernandez:2013hea,Hernandez:2014vta,CarcamoHernandez:2010im,Bhattacharyya:2012pi,Campos:2014lla}
and better than, for example, those in Refs.~\cite{Chen:2007afa,Xing:2010iu,Branco:2011wz,CarcamoHernandez:2012xy,CarcamoHernandez:2012xy,Vien:2014ica,Abbas:2014ewa,Ishimori:2014jwa,Ishimori:2014nxa}%
.

\section{Lepton masses and mixings}

\label{leptonsector} This warped $S_{3}$ flavor model generates the viable
and predictive quark textures proposed in \cite{Hernandez:2014zsa} as shown
in section \ref{quarksector}. We now proceed to analyze the lepton sector of
the model. From the charged lepton Yukawa terms of Eq. (\ref{Lyl}) it
follows that the charged lepton mass matrix takes the following form: 
\begin{equation}
M_{l}=\left( 
\begin{array}{ccc}
\varepsilon _{11}^{\left( l\right) }\frac{v_{\xi }v_{\eta }}{M_{5}^{2}} & 0
& 0 \\ 
0 & \varepsilon _{22}^{\left( l\right) }\frac{v_{\xi }}{M_{5}} & \varepsilon
_{23}^{\left( l\right) } \\ 
0 & \varepsilon _{32}^{\left( l\right) }\frac{v_{\xi }}{M_{5}} & \varepsilon
_{33}^{\left( l\right) }%
\end{array}%
\right) \frac{v}{\sqrt{2}}=\left( 
\begin{array}{ccc}
w & 0 & 0 \\ 
0 & r & s \\ 
0 & x & z%
\end{array}%
\right) \frac{v}{\sqrt{2}},
\end{equation}%
where: 
\begin{eqnarray}
\varepsilon _{nn}^{\left( l\right) } &=&\lambda _{nn}^{\left( l\right) }%
\frac{k}{M_{5}^{\frac{3}{2}}}\sqrt{\frac{2k\left( 1\pm \sqrt{4+a}\right)
\left( 1-2y_{nL}^{\left( l\right) }\right) \left( 1-2y_{nR}^{\left( l\right)
}\right) }{\left[ e^{2\left( 1\pm \sqrt{4+a}\right) k\pi R}-1\right] \left[
e^{\left( 1-2y_{nL}^{\left( l\right) }\right) k\pi R}-1\right] \left[
e^{\left( 1-2y_{nR}^{\left( l\right) }\right) k\pi R}-1\right] }},\hspace{1cm%
}n=1,2,  \notag \\
\varepsilon _{23}^{\left( l\right) } &=&\lambda _{23}^{\left( l\right) }%
\frac{k}{M_{5}^{\frac{3}{2}}}\sqrt{\frac{2k\left( 1\pm \sqrt{4+a}\right)
\left( 1-2y_{2L}^{\left( l\right) }\right) \left( 1-2y_{3R}^{\left( l\right)
}\right) }{\left[ e^{2\left( 1\pm \sqrt{4+a}\right) k\pi R}-1\right] \left[
e^{\left( 1-2y_{2L}^{\left( l\right) }\right) k\pi R}-1\right] \left[
e^{\left( 1-2y_{3R}^{\left( l\right) }\right) k\pi R}-1\right] }}\frac{%
e^{\left( 2-y_{3L}^{\left( l\right) }-y_{2R}^{\left( l\right) }\pm \sqrt{4+a}%
\right) k\pi R}-1}{\left[ 2-y_{2L}^{\left( l\right) }-y_{3R}^{\left(
l\right) }\pm \sqrt{4+a}\right] },  \notag \\
\varepsilon _{33}^{\left( l\right) } &=&\lambda _{33}^{\left( l\right) }%
\frac{1}{M_{5}^{\frac{1}{2}}}\sqrt{\frac{2k\left( 1\pm \sqrt{4+a}\right)
\left( 1-2y_{3L}^{\left( l\right) }\right) \left( 1-2y_{3R}^{\left( l\right)
}\right) }{\left[ e^{2\left( 1\pm \sqrt{4+a}\right) k\pi R}-1\right] \left[
e^{\left( 1-2y_{3L}^{\left( l\right) }\right) k\pi R}-1\right] \left[
e^{\left( 1-2y_{3R}^{\left( l\right) }\right) k\pi R}-1\right] }}\frac{%
e^{\left( 2-y_{3L}^{\left( l\right) }-y_{3R}^{\left( l\right) }\pm \sqrt{4+a}%
\right) k\pi R}-1}{\left[ 2-y_{3L}^{\left( l\right) }-y_{3R}^{\left(
l\right) }\pm \sqrt{4+a}\right] },  \notag \\
\varepsilon _{32}^{\left( l\right) } &=&\lambda _{32}^{\left( l\right) }%
\frac{1}{M_{5}^{\frac{1}{2}}}\sqrt{\frac{2k\left( 1\pm \sqrt{4+a}\right)
\left( 1-2y_{3L}^{\left( l\right) }\right) \left( 1-2y_{2R}^{\left( l\right)
}\right) }{\left[ e^{2\left( 1\pm \sqrt{4+a}\right) k\pi R}-1\right] \left[
e^{\left( 1-2y_{3L}^{\left( l\right) }\right) k\pi R}-1\right] \left[
e^{\left( 1-2y_{2R}^{\left( l\right) }\right) k\pi R}-1\right] }},
\end{eqnarray}%
where $y_{1R}^{\left( l\right) }=y_{2R}^{\left( l\right) }=y_{R}^{\left(
l\right) }$ as follows from the fact thal $l_{1R}$ and $l_{2R}$ are unified
into a $S_{3}$ doublet as seen from Eq. (\ref{Fermionassignments}).

Therefore, $M_{l}M_{l}^{T}$ can be approximately diagonalized by a rotation
matrix $R_{l}$ according to: 
\begin{equation}
R_{l}^{T}M_{l}M_{l}^{T}R_{l}=\left( 
\begin{array}{ccc}
m_{e}^{2} & 0 & 0 \\ 
0 & m_{\mu }^{2} & 0 \\ 
0 & 0 & m_{\tau }^{2}%
\end{array}%
\right) ,\,\quad \quad \quad R_{l}=\left( 
\begin{array}{ccc}
1 & 0 & 0 \\ 
0 & \cos \gamma & -\sin \gamma \\ 
0 & \sin \gamma & \cos \gamma%
\end{array}%
\right) \allowbreak ,\,\quad \quad \quad \tan 2\gamma =\frac{2\left(
rx+sz\right) }{\left( r^{2}+s^{2}\right) -\left( x^{2}+z^{2}\right) }.
\end{equation}

The masses for charged leptons take the form: 
\begin{eqnarray}
m_{e} &=&\varepsilon _{11}^{\left( l\right) }\frac{v_{\xi }v}{\sqrt{2}M_{5}}%
,\,\quad \quad \quad  \notag \\
m_{\mu } &=&\sqrt{\frac{1}{2}\left( r^{2}+s^{2}+x^{2}+z^{2}\right) -\frac{1}{%
2}\sqrt{\left( r^{2}+2rz+s^{2}-2sx+x^{2}+z^{2}\right) \left(
r^{2}-2rz+s^{2}+2sx+x^{2}+z^{2}\right) }}\frac{v}{\sqrt{2}},\,\quad \quad
\quad  \notag \\
m_{\tau } &=&\sqrt{\frac{1}{2}\left( r^{2}+s^{2}+x^{2}+z^{2}\right) +\frac{1%
}{2}\sqrt{\left( r^{2}+2rz+s^{2}-2sx+x^{2}+z^{2}\right) \left(
r^{2}-2rz+s^{2}+2sx+x^{2}+z^{2}\right) }}\frac{v}{\sqrt{2}}.  \label{mlc}
\end{eqnarray}

To show that the charged lepton texture can fit the experimental data and in
order to simplify our analysis, we adopt a benchmark where we set:

\begin{equation}
y_{1L}^{\left( l\right) }=y_{2L}^{\left( l\right) }=y_{3L}^{\left( l\right)
}=y_{L}^{\left( l\right) },\hspace{1cm}\hspace{1cm}\lambda _{23}^{\left(
l\right) }=\lambda _{33}^{\left( l\right) }=\lambda ^{\left( l\right) },%
\hspace{1cm}\hspace{1cm}\lambda _{32}^{\left( l\right) }=1.
\end{equation}

We assume equality of left handed leptonic profiles to avoid a huge
hierarchy in the entries of the neutrino mass matrix, as required by the
neutrino oscillation experimental data which favors a moderate hierarchy
among these entries. Our two assumptions concerning the charged lepton
Yukawa couplings are motivated by the partial universality of these
couplings and by naturalness arguments. Let us note that the charged lepton
mass hierarchy is caused by the different localizations of the right handed
charged leptonic fields $l_{R}$ and $l_{3R}$ in the extra dimension as well
as by the different power dependances in terms of the Wolfenstein parameter $%
\lambda =0.225$ presented in the different columns of the charged lepton
mass matrix. Consequently it is natural to assume partial universality in
the charged lepton Yukawa couplings.

In addition, the $S_{3}$ discrete symmetry leads to the following constraint:

\begin{equation}
y_{1R}^{\left( l\right) }=y_{2R}^{\left( l\right) }=y_{R}^{\left( l\right) }.
\end{equation}

Then, we proceed to adjust the parameters $y_{R}^{\left( l\right) }$, $%
y_{L}^{\left( l\right) }$, $y_{3R}^{\left( l\right) }$, $\lambda ^{\left(
l\right) }$, $\lambda _{11}^{\left( l\right) }$ and $\lambda _{22}^{\left(
l\right) }$ to reproduce the experimental values of the charged lepton
masses finding the following best fit result:

\begin{eqnarray}
\lambda _{11}^{\left( l\right) } &=&3.147,\hspace{1cm}\hspace{1cm}\lambda
_{22}^{\left( l\right) }=2.218,\hspace{1cm}\hspace{1cm}\lambda ^{\left(
l\right) }=0.484,  \notag \\
y_{L}^{\left( l\right) } &=&7.323713,\hspace{1cm}y_{R}^{\left( l\right)
}=0.5207487,\hspace{1cm}y_{3R}^{\left( l\right) }=0.501.
\end{eqnarray}

The full $5\times 5$ neutrino mass matrix arises from the the neutrino
Yukawa interactions given in Eq. (\ref{Lyl}) and is given by: 
\begin{equation}
M_{\nu }=\left( 
\begin{array}{cc}
0_{3\times 3} & M_{\nu }^{D} \\ 
\left( M_{\nu }^{D}\right) ^{T} & M_{R}%
\end{array}%
\right) ,  \label{Mnu}
\end{equation}%
where: 
\begin{eqnarray}
M_{\nu }^{D} &=&\left( 
\begin{array}{cc}
0 & \varepsilon _{12}^{\left( \nu \right) }\frac{v_{\eta }}{M_{5}} \\ 
\varepsilon _{21}^{\left( \nu \right) } & \varepsilon _{22}^{\left( \nu
\right) } \\ 
\varepsilon _{31}^{\left( \nu \right) } & \varepsilon _{32}^{\left( \nu
\right) }%
\end{array}%
\right) \frac{v}{\sqrt{2}}=\left( 
\begin{array}{cc}
0 & f \\ 
b & c \\ 
h & d%
\end{array}%
\right) \frac{v}{\sqrt{2}},\hspace{1cm}M_{R}=\left( 
\begin{array}{cc}
M_{1\nu } & M_{3\nu } \\ 
M_{3\nu } & M_{2\nu }%
\end{array}%
\right) ,  \notag \\
M_{R}^{-1} &=&\left( 
\begin{array}{cc}
\frac{M_{2\nu }}{M_{\nu }M_{2\nu }-M_{3\nu }^{2}} & -\frac{M_{3\nu }}{M_{\nu
}M_{2\nu }-M_{3\nu }^{2}} \\ 
-\frac{M_{3\nu }}{M_{\nu }M_{2\nu }-M_{3\nu }^{2}} & \frac{M_{\nu }}{M_{\nu
}M_{2\nu }-M_{3\nu }^{2}}%
\end{array}%
\right) \allowbreak =\left( 
\begin{array}{cc}
p & r \\ 
r & q%
\end{array}%
\right) ,
\end{eqnarray}

\begin{eqnarray}
\varepsilon _{n1}^{\left( \nu \right) } &=&\lambda _{n1}^{\left( \nu \right)
}\frac{k}{M_{5}}\sqrt{\frac{2\left( 1\pm \sqrt{4+a}\right) \left(
1-2y_{nL}^{\left( \nu \right) }\right) }{\left[ e^{2\left( 1\pm \sqrt{4+a}%
\right) k\pi R}-1\right] \left[ e^{\left( 1-2y_{nL}^{\left( \nu \right)
}\right) k\pi R}-1\right] }},\hspace{1cm}n=2,3,  \notag \\
\varepsilon _{j2}^{\left( \nu \right) } &=&\lambda _{j2}^{\left( \nu \right)
}\frac{k}{M_{5}}\sqrt{\frac{2\left( 1\pm \sqrt{4+a}\right) \left(
1-2y_{jL}^{\left( \nu \right) }\right) }{\left[ e^{2\left( 1\pm \sqrt{4+a}%
\right) k\pi R}-1\right] \left[ e^{\left( 1-2y_{jL}^{\left( \nu \right)
}\right) k\pi R}-1\right] }},\hspace{1cm}j=1,2,3.
\end{eqnarray}

As the Majorana neutrino masses are much larger than the electroweak
symmetry breaking scale $v=256$ GeV, i.e., $\left( M_{R}\right) _{ii}\gg v$,
active neutrinos acquire small masses via a type-I seesaw mechanism mediated
by the Majorana neutrinos. Then, the light neutrino mass matrix reads 
\begin{eqnarray}
M_{L} &=&M_{\nu }^{D}M_{R}^{-1}\left( M_{\nu }^{D}\right) ^{T}=\left( 
\begin{array}{cc}
0 & f \\ 
b & c \\ 
h & d%
\end{array}%
\right) \left( 
\begin{array}{cc}
p & r \\ 
r & q%
\end{array}%
\right) \left( 
\begin{array}{ccc}
0 & b & h \\ 
f & c & d%
\end{array}%
\right) \frac{v^{2}}{2}  \notag \\
&=&\left( 
\begin{array}{ccc}
f^{2}q & f\left( br+cq\right)  & f\left( dq+hr\right)  \\ 
f\left( br+cq\right)  & pb^{2}+2rbc+qc^{2} & b\left( dr+hp\right) +c\left(
dq+hr\right)  \\ 
f\left( dq+hr\right)  & b\left( dr+hp\right) +c\left( dq+hr\right)  & 
qd^{2}+2rdh+ph^{2}%
\end{array}%
\right) \frac{v^{2}}{2}\allowbreak 
\end{eqnarray}

In order to show that the light neutrino mass matrix given above is
consistent with the experimental data on neutrino oscillations and in order
to simplify our analysis, we adopt a benchmark where we set:%
\begin{equation}
p=q,\hspace{1cm}h=-d,\hspace{1cm}b=c.
\end{equation}
so that the light active neutrino mass matrix takes the form:
\begin{equation}
M_{L}=\left( 
\begin{array}{ccc}
f^{2}p & bf\left( p+r\right)  & df\left( p-r\right)  \\ 
bf\left( p+r\right)  & 2b^{2}\left( p+r\right)  & 0 \\ 
df\left( p-r\right)  & 0 & 2d\left( dp-dr\right) 
\end{array}%
\right) \allowbreak \allowbreak \frac{v^{2}}{2}
\end{equation}
which implies that the light active neutrino mass matrix does not contribute to $\sin^{2}\theta _{23}$. The leptonic mixing parameter $\sin ^{2}\theta _{23}$
solely arises from the charged lepton sector and is found to be equal to $0.507$, which is consistent with its corresponding experimental value within
the $1\sigma $ experimentally allowed range. Varying the parameters $p$, $r$%
, $b$, $d$\ and $f$ we fitted $\Delta m_{21}^{2}$, $\Delta m_{31}^{2}$\
(note that we define $\Delta m_{ij}^{2}=m_{i}^{2}-m_{j}^{2}$), $\sin
^{2}\theta _{12}$ and $\sin ^{2}\theta _{13}$ to the experimental values 
\cite{Forero:2014bxa} in Table \ref{NH} for the normal hierarchy neutrino
mass spectrum. The best fit result is: 
\begin{eqnarray}
\Delta m_{21}^{2} &=&7.55\times 10^{-5}\mbox{eV}^{2},\quad \quad \quad
\Delta m_{31}^{2}=2.51\times 10^{-3}\mbox{eV}^{2},  \notag \\
m_{\nu _{1}} &=&0,\,\quad \quad \quad m_{\nu _{2}}\simeq 9\mbox{meV},\quad
\quad \quad \,m_{\nu _{3}}\simeq 50\mbox{meV},\,\quad \quad \quad   \notag \\
\sin ^{2}\theta _{12} &=&0.311,\quad \quad \quad \sin ^{2}\theta
_{13}=0.024,\quad \quad \quad \sin ^{2}\theta _{23}=0.507,  \notag \\
\quad b &\simeq &0.38,\quad \quad \quad f\simeq 0.52,\quad \quad \quad
d\simeq 1.57,\quad \quad \quad   \label{Parameterfit-NH} \\
q &\simeq &-1.72\times 10^{-16}\mbox{GeV}^{-1},\quad \quad \quad r\simeq
-4.98\times 10^{-16}\mbox{GeV}^{-1}.  \notag
\end{eqnarray}

\begin{table}[tbh]
\begin{tabular}{|c|c|c|c|c|c|}
\hline
Parameter & $\Delta m_{21}^{2}$($10^{-5}$eV$^2$) & $\Delta m_{31}^{2}$($%
10^{-3}$eV$^2$) & $\left( \sin ^{2}\theta _{12}\right) _{\exp }$ & $\left(
\sin ^{2}\theta _{23}\right) _{\exp }$ & $\left( \sin ^{2}\theta
_{13}\right) _{\exp }$ \\ \hline
Best fit & $7.60$ & $2.48$ & $0.323$ & $0.567$ & $0.0234$ \\ \hline
$1\sigma $ range & $7.42-7.79$ & $2.41-2.53$ & $0.307-0.339$ & $0.439-0.599$
& $0.0214-0.0254$ \\ \hline
$2\sigma $ range & $7.26-7.99$ & $2.35-2.59$ & $0.292-0.357$ & $0.413-0.623$
& $0.0195-0.0274$ \\ \hline
$3\sigma $ range & $7.11-8.11$ & $2.30-2.65$ & $0.278-0.375$ & $0.392-0.643$
& $0.0183-0.0297$ \\ \hline
\end{tabular}%
\caption{Experimental ranges of neutrino squared mass differences and
leptonic mixing angles, from Ref. \protect\cite{Forero:2014bxa}, for the
normal hierarchy neutrino mass spectrum.}
\label{NH}
\end{table}

\begin{table}[tbh]
\begin{center}
\begin{tabular}{c|l|l}
\hline\hline
Observable & Model value & Experimental value \\ \hline
$m_{e}(MeV)$ & \quad $0.487$ & \quad $0.487$ \\ \hline
$m_{\mu }(MeV)$ & \quad $103$ & \quad $102.8\pm 0.0003$ \\ \hline
$m_{\tau }(GeV)$ & \quad $1.75$ & \quad $1.75\pm 0.0003$ \\ \hline
$\Delta m_{21}^{2}$($10^{-5}$eV$^{2}$) & \quad $7.60$ & \quad $7.60_{-0.18}^{+0.19}$ \\ \hline
$\Delta m_{31}^{2}$($10^{-3}$eV$^{2}$) & \quad $2.48$ & \quad $2.48_{-0.07}^{+0.05}$ \\ \hline
$\sin ^{2}\theta _{12}$ & \quad $0.311$ & \quad $0.323\pm 0.016$ \\ \hline
$\sin ^{2}\theta _{23}$ & \quad $0.507$ & \quad $0.567_{-0.128}^{+0.032} $
\\ \hline
$\sin ^{2}\theta _{13}$ & \quad $0.024$ & \quad $0.0234\pm 0.0020$ \\ \hline
\end{tabular}%
\end{center}
\par
\caption{Model and experimental values of the charged lepton masses, neutrino mass squared splittings and leptonic mixing parameters for the normal hierarchy neutrino mass spectrum.}
\label{Observables0}
\end{table}
In Table \ref{Observables0} we show the model and experimental values for the physical observables of the lepton sector for the normal hierarchy neutrino mass spectrum. Comparing Eq (\ref{Parameterfit-NH}) with Table \ref{NH} we see that the mass squared splittings $\Delta m_{21}^{2}$ and $\Delta m_{31}^{2}$ and mixing parameters $\sin ^{2}\theta _{12}$, $\sin^{2}\theta_{13}$ and $\sin^{2}\theta _{23}$ are in excellent agreement with the experimental data as they are inside the $1\sigma $ experimentally allowed range. Note that here we considered all leptonic parameters to be real for simplicity, but a non-vanishing CP phase in the PMNS mixing matrix can be generated by making complex any of the off diagonal charged lepton Yukawa couplings, e.g. $\lambda _{23}^{(l)}$ or $\lambda _{32}^{(l)}$. It is noteworthy that, in view of the parametric freedom of our model it is always possible to fit any physical observable of both quark and lepton sector independently of its degree of precision. Thus, for example an increase in the precision of experimental value of the leptonic mixing parameter $sin^{2}\theta _{13}$ will not rule out our model

Now we can predict the amplitude for neutrinoless double beta ($0\nu \beta
\beta $) decay, which is proportional to the effective Majorana neutrino
mass 
\begin{equation}
m_{\beta \beta }=\left\vert \sum_{k}U_{ek}^{2}m_{\nu _{k}}\right\vert ,
\label{mee}
\end{equation}%
where $U_{ek}^{2}$ and $m_{\nu _{k}}$ are the PMNS mixing matrix elements
and the Majorana neutrino masses, respectively.

Then, we predict the following effective neutrino mass:

\begin{equation}
m_{\beta \beta }\approx 1.4\ \mbox{meV}.  \label{eff-mass-pred}
\end{equation}

Our value for the effective neutrino mass is beyond the reach of the present
and forthcoming neutrinoless double beta decay experiments. The current best
upper bound for the effectice neutrino mass, i.e., $m_{\beta \beta }\leq 160$
meV arises from the EXO-200 experiment \cite{Auger:2012ar} $T_{1/2}^{0\nu
\beta \beta }(^{136}\mathrm{Xe})\geq 1.6\times 10^{25}$ yr at the 90 \% CL.
An improvement of this upper bound is expected within the not too far
future. The GERDA experiment \cite{Abt:2004yk,Ackermann:2012xja} is
upgrading to \textquotedblleft phase-II\textquotedblright , and is expected
to reach 
\mbox{$T^{0\nu\beta\beta}_{1/2}(^{76}{\rm Ge}) \geq 2\times
10^{26}$ yr}, corresponding to $m_{\beta \beta }\leq 100$ MeV. A bolometric
CUORE experiment, using ${}^{130}\mbox{Te}$ \cite{Alessandria:2011rc}, is
under construction. Its sensitivity is estimated at about $T_{1/2}^{0\nu
\beta \beta }(^{130}\mathrm{Te})\sim 10^{26}$ yr, which corresponds to %
\mbox{$m_{\beta\beta}\leq 50$ meV.} There are several proposals for
ton-scale future $0\nu \beta \beta $ experiments with $^{136}$Xe \cite%
{KamLANDZen:2012aa,Albert:2014fya} and $^{76}$Ge \cite%
{Abt:2004yk,Guiseppe:2011me} estimating sensitivities over $T_{1/2}^{0\nu
\beta \beta }\sim 10^{27}$ yr, corresponding to $m_{\beta \beta }\sim 12-30$
meV. Recent experimental reviews can be found in Ref. \cite{Bilenky:2014uka}
and references therein. Consequently, as indicated by Eq. (\ref%
{eff-mass-pred}) we get a prediction of $T_{1/2}^{0\nu \beta \beta }$, which
is at the level of sensitivities of the next generation or next-to-next
generation $0\nu \beta \beta $ experiments.

\section{Conclusions}

\label{conclusions}

In this paper we proposed an extension of the Standard Model where a warped
extra dimension is supplemented by the discrete flavor symmetry $S_{3}\times
Z_{2}\times Z_{4}$ and the particle content is extended by including two
heavy right handed Majorana neutrinos and scalar fields responsible for the
breaking of the discrete symmetry. To the best of our knowledge, our model
is the first implementation of the $S_{3}$ flavor symmetry in a five
dimensional warped framework. We examined a particular choice of $S_{3}$
assignments in the quark sector, which leads to a mixing inspired texture
where the down-type quark sector contributes to the Cabbibo mixing, whereas
the up-type quark sector contributes to the remaining mixing angles. In the
lepton sector, the effective neutrino masses are obtained through a type I
seesaw with two right-handed neutrinos, and the charged lepton mass texture
contributes to the mixing in the 2-3 plane. The flavor symmetry is
responsible for several texture zeros in the mass matrices of each sector.
The model is capable of a very good fit to the fermion data, and the mass
hierarchy and mixing angles are explained by a combination of the discrete
symmetries and the exponential suppressions arising from the warped
spacetime.

\section{Acknowledgments}

This project has received funding from the European Union's Seventh
Framework Programme for research, technological development and
demonstration under grant agreement no PIEF-GA-2012-327195 SIFT. A.E.C.H
thanks Southampton University for hospitality where part of this work was
done. A.E.C.H was supported by Fondecyt (Chile), Grant No. 11130115 and by
DGIP internal Grant No. 111458.

\appendix

\section{The product rules for $S_{3}$.}

\label{S3}

The $S_{3}$ discrete group contains 3 irreducible representations: $\mathbf{1%
}$, $\mathbf{1}^{\prime }$ and $\mathbf{2}$. Considering $\left(
x_{1},x_{2}\right) ^{T}$\ and $\left( y_{1},y_{2}\right) ^{T}$ as the basis
vectors for two $S_{3}$ doublets and $y%
{\acute{}}%
$ an $S_{3}$ non trivial singlet, the multiplication rules of the $S_{3}$
group for the case of real representations take the form \cite%
{Ishimori:2010au}: 
\begin{equation}
\left( 
\begin{array}{c}
x_{1} \\ 
x_{2}%
\end{array}%
\right) _{\mathbf{2}}\otimes \left( 
\begin{array}{c}
y_{1} \\ 
y_{2}%
\end{array}%
\right) _{\mathbf{2}}=\left( x_{1}y_{1}+x_{2}y_{2}\right) _{\mathbf{1}%
}+\left( x_{1}y_{2}-x_{2}y_{1}\right) _{\mathbf{1}^{\prime }}+\left( 
\begin{array}{c}
x_{2}y_{2}-x_{1}y_{1} \\ 
x_{1}y_{2}+x_{2}y_{1}%
\end{array}%
\right) _{\mathbf{2}},  \label{6}
\end{equation}%
\begin{equation}
\left( 
\begin{array}{c}
x_{1} \\ 
x_{2}%
\end{array}%
\right) _{\mathbf{2}}\otimes \left( y%
{\acute{}}%
\right) _{\mathbf{1}^{\prime }}=\left( 
\begin{array}{c}
-x_{2}y%
{\acute{}}
\\ 
x_{1}y%
{\acute{}}%
\end{array}%
\right) _{\mathbf{2}},\hspace{1cm}\hspace{1cm}\left( x%
{\acute{}}%
\right) _{\mathbf{1}^{\prime }}\otimes \left( y%
{\acute{}}%
\right) _{\mathbf{1}^{\prime }}=\left( x%
{\acute{}}%
y%
{\acute{}}%
\right) _{\mathbf{1}}.  \label{7}
\end{equation}

\end{document}